# Charge and Energy Transfer Dynamics of Hybridized Exciton-Polaritons in 2D Halide Perovskites

*Surendra B. Anantharaman[1,\*], Jason Lynch[1], Christopher E. Stevens[2,3], Christopher Munley[4], Chentao Li[5], Jin Hou[6,7], Hao Zhang[6,8], Andrew Torma[6,8], Thomas Darlington[9], Francis Coen[1], Kevin Li[1], Arka Majumdar[4,10], P. James Schuck[9], Aditya Mohite[5,6], Hayk Harutyunyan[5], Joshua R. Hendrickson[3], Deep Jariwala[1,\*]*

[1] Department of Electrical and Systems Engineering, University of Pennsylvania, Philadelphia, Pennsylvania 19104, United States

[2] KBR Inc., Beavercreek, Ohio, 45431, United States

[3] Air Force Research Laboratory, Sensors Directorate, Wright-Patterson Air Force Base, Ohio 45433, United States

[4] Department of Physics, University of Washington, Seattle, Washington 98195, United States

[5] Department of Physics, Emory University, Atlanta, Georgia 30322, United States

[6] Department of Chemical and Biomolecular Engineering, Rice University, Houston, Texas 77005, United States

[7] Department of Materials Science and Nanoengineering, Rice University, Houston, Texas 77005, United States

[8] Applied Physics Program, Smalley-Curl Institute, Rice University, Houston, Texas 77005, United states

[9] Department of Mechanical Engineering, Columbia University, New York, New York 10027, United States

[10] Department of Electrical and Computer Engineering, University of Washington, Seattle, Washington 98195, United States

Corresponding authors: surendra.anantharaman@gmail.com, dmj@seas.upenn.edu

## Abstract

Excitons, bound electron-hole pairs, in Two-Dimensional Hybrid Organic Inorganic Perovskites (2D HOIPs) are capable of forming hybrid light-matter states known as exciton-polaritons (E-Ps) when the excitonic medium is confined in an optical cavity. In the case of 2D HOIPs, they can self-hybridize into E-Ps at specific thicknesses of the HOIP crystals that form a resonant optical cavity with the excitons. However, the fundamental properties of these self-hybridized E-Ps in 2D HOIPs, including their role in ultrafast energy and/or charge transfer at interfaces, remain unclear. Here, we demonstrate that > 0.5 µm thick 2D HOIP crystals on Au substrates are capable of supporting multiple-orders of self-hybridized E-P modes. These E-Ps have high Q factors (> 100) and modulate the optical dispersion for the crystal to enhance sub-gap absorption and emission. Through varying excitation energy and ultrafast measurements, we also confirm energy transfer from higher energy upper E-Ps to lower energy, lower E-Ps. Finally, we also demonstrate that E-Ps are capable of charge transport and transfer at interfaces. Our findings provide new insights into charge and energy transfer in E-Ps opening new opportunities towards their manipulation for polaritonic devices.

**Introduction**

Halide perovskites, a class of hybrid organic-inorganic semiconductors, have garnered considerable attention for their potential applications in optoelectronic devices including solar cells, light emitting diodes, and lasers[1,2]. These materials range in dimension from 0D quantum dots to 3D bulk crystals, each with varying physical properties as well as their own strengths and weaknesses with regards to device performance[3,4]. For example, 3D perovskites demonstrate high solar cell performance, but are also highly susceptible to device degradation. Recent research has addressed this challenge facing 3D perovskites, finding enhanced stability in integrated heterostructures of 2D and 3D perovskites[5–8]. However, these 2D perovskites have much stronger exciton binding energy than their 3D counterparts due to quantum confinement effects, which reduces the probability of electron-hole pairs disassociating at charge-separation interfaces which also results in a high emission quantum yield[9–11]. Therefore, the 2D pervoskites serve as a near-ideal system for investigatiing strong-coupling of self-hybridized light and matter states (e.g. exciton-polaritons) and their fundamental properties in an open-cavity[12,13]. In particular, the extent to which self-hybridized exciton-polaritons can modify the optical dispersion of a crystalline host is an open question. Further, how self-hybridized exciton-polaritons (E-Ps) transfer charge or energy between the various cascade of states or to a nearby electronically coupled layer is also unknown.

Strong light-matter interactions lead to part-light and part-matter quasiparticle states called E-Ps that modify the optical dispersion of the system, and enable tunable emission that depends on the detuning between the uncoupled excitonic and photonic cavity states[14,15]. Although these states are typically observed when the semiconductors are placed inside a dielectric cavity[16–20] or on a plasmonic interface[21–

[23], 2D perovskites have shown self-hybridized E-Ps in the absence of an external cavity[16,24]. In the presence of multiple internal cavity modes, multiple polariton branches form below the band gap of the 2D perovskite. Each of these branches alters the relaxation pathway of E-Ps. Therefore, a proper understanding of these multiple polariton branch systems is necessary to understand their effects on charge separation for photovoltaics and on Bose-Einstein condensate formation for polaritonic lasing.

In this work, we answer three important fundamental questions about the optical properties emerging from strong light-matter coupling in 2D perovskites. Can E-Ps modify the optical dispersion of a 2D perovskite crystal? Can these self-hybridized E-Ps undergo polariton condensation and/or lasing in an open cavity system? And finally, is it possible to transduce energy from E-Ps at energies below the primary exciton resonance? We find that with an increase of the perovskite flake thickness to >500 nm, strong light-matter coupling leads to the formation of multiple, self-hybridized E-P branches in the optical dispersion. We use transfer-matrix calculations to understand the contribution of the photonic interaction, and we perform detailed investigations via photoluminescence (PL) excitation spectroscopy, temperature-dependent PL, PL mapping, pump-probe spectroscopy, and time-resolved PL measurements to understand the origin of sub-bandgap absorption and emission states. These measurements conclusively prove the existence of E-Ps that modify the optical dispersion of the 2D perovskite layers in both absorption and emission. We further find these polariton states explain the peak splitting at the excitonic energy, and the emergence of multiple emission peaks below the bandgap resulting from sub-bandgap excitation. Importantly, we also find that there is energy transfer/funneling from the upper polariton to the lower polariton branches. This energy transfer into multiple, lower polariton states prevents condensation and lasing from polaritons at pump powers below the damage threshold of the perovskite crystals. Finally, we also use a perovskite/few-layer graphene van der Waals heterostructure to prove that these sub-bandgap polariton states can, in principle, be electrically harvested and thereby demonstrate the feasibility of introducing strong light-matter coupling in energy harvesting applications such as polaritonic photodetectors and photovoltaics.

**Results and Discussion**

We exfoliate micro-scale crystals of 2D Ruddlesden-Popper (RP) perovskites from ~mm size bulk crystals grown in a solution as described in earlier works[25–29]. A schematic of the exfoliated 2D perovskite (($BA)_2(CH_3NH_3)_{n-1}Pb_nI_{3n+1}$ (n=1), hereafter RP1, n=2 as RP2 and n=3 as RP3) with different flake thicknesses is shown in Figure 1a. Photoluminescence signals from spontaneous emission can be observed in hybrid organic-inorganic perovskites (HOIPs) thin films of thickness less than 10 nm on an Au substrate. In our previous work, we reported that an increase of film thickness to 100 nm is accompanied by the formation of single-mode, self-hybridized exciton-polaritons in an open cavity[24]. In the present study, we explore photonic bands emerging in flakes of thickness >500 nm. This increase in thickness induces higher-order open cavities within the flake. As a result, the excitons interact with multiple cavity modes and hybridize to form multiple E-P modes. In Figure 1b, we schematically elucidate the relationship between increased perovskite flake thickness and emergence of multiple polariton branches. A critical flake thickness of 100 nm gives way to first-order E-P which form upper and lower polariton branches, denoted as UPB1 and LPB1, respectively. Further increases in thickness beyond 100 nm introduce higher-order cavity modes, which interact with the exciton to form UPB2 and LPB2.

Experimentally, we have investigated reflectance and PL spectroscopy studies from the RP1 perovskite exfoliated on an Au substrate (Au/RP1) as shown in Figure 1c. The reflectance data shows multiple dips below the bandgap edge. These dips correspond to the E-P states formed from strong light-matter coupling as reported elsewhere[5]. Upon optically pumping with a 405 nm laser, the emission from all of the states shows a close correlation with the reflectance spectrum. The appearance of multiple peaks in the PL shows that the excited carriers can relax to any of the lower polariton branches before returning directly to its ground state and emitting a photon. When the sample is optically pumped with a 633 nm laser there is small absorption since the exciton for RP1 is at 509 nm, as seen in the reflectance spectrum. Therefore, the weak PL peak around 680 nm that was observed is due to sub-bandgap absorption and emission.

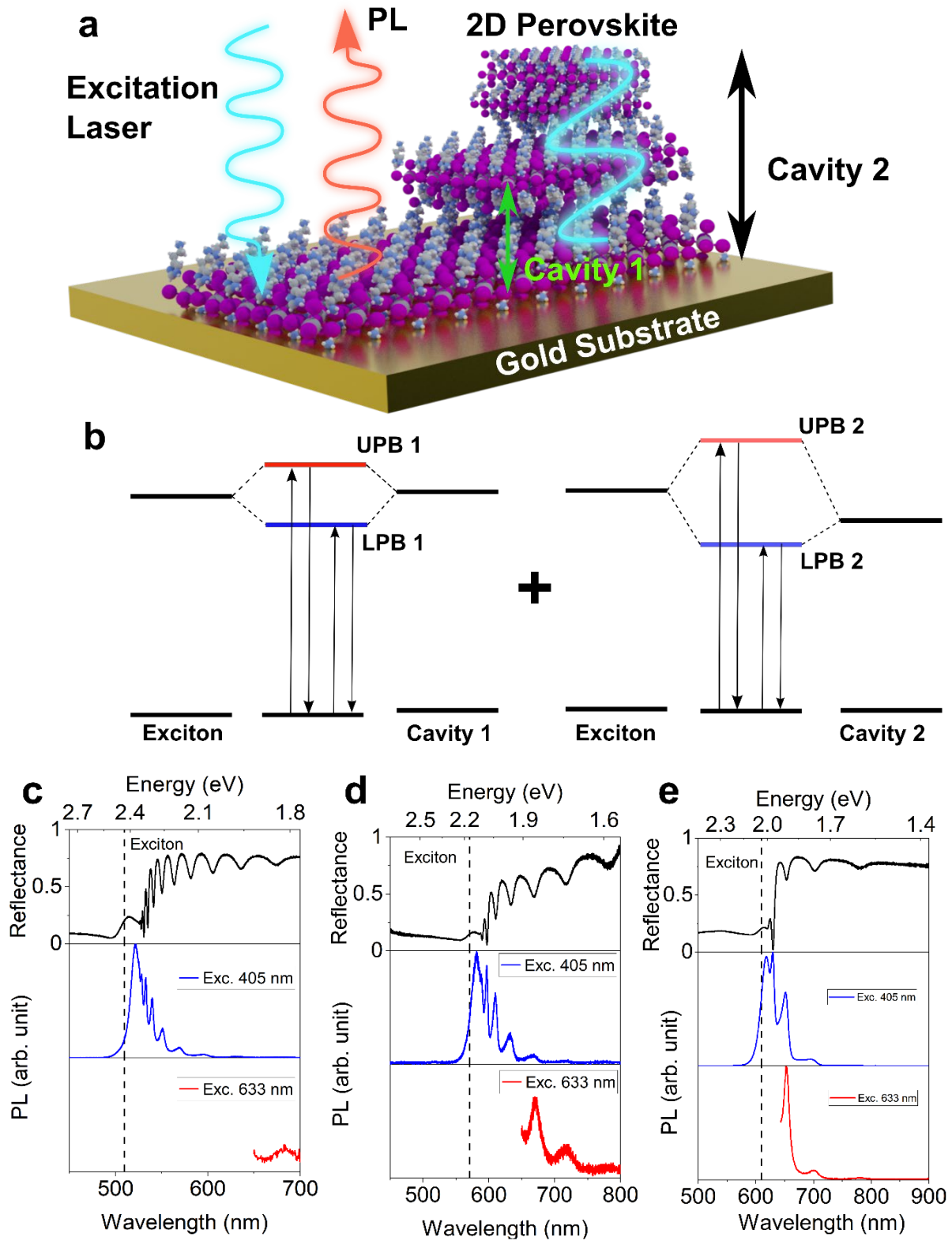

**Figure 1. Sub-bandgap photon absorption and emission in RP1 and RP2 on Au.**
(a) Schematic showing the origin of the sub-bandgap photon absorption states upon strong light-matter interaction in a self-hybridized multi-cavity mode. (b) Energy band (UPB1, LPB1, UPB2, LPB2) formation upon strong light-matter interaction with multiple cavity modes (cavity 1 and cavity 2) from the same exciton energy of the semiconductor (i.e. 2D perovskites in this study). Sub-band gap absorption and emission transitions are shown as arrows. Experimental observation of sub-bandgap absorption (or reflectance) and emission from (c) RP1, (d) RP2, and (e) RP3 flakes exfoliated on the Au substrate. Emission states from the hybridized system recorded using 633 excitation laser match exactly with the emission upon excitation with 405 nm laser and a corresponding reflectance state.

To further evaluate the role of strong light-matter coupling in other perovskites, we investigated RP2 and RP3 on Au substrates as well. For Au/RP2, reflectance spectroscopy reveals similar multiple polariton branches as were observed in Au/RP1, as shown in Figure 1d. These reflectance states are in agreement with PL states at room temperature, including the 580 nm excitonic peak observed under excitation from a 405 nm laser, as reported in our previous work. In these self-hybridized RP2 systems, we observe multiple orders of emission from exciton-polaritons, confirming the existence of the UPB, LPB, and higher order (HO) modes. The HO modes can exhibit large Q factors (132 in RP1) which are extremely sensitive to the surrounding dielectric medium such as polymers and can be potentially used for colorimetric sensing (see supporting information Figures S1 and S2). In addition, multiple LPB (LPB1, LPB2 etc.) can be observed upon optical pumping with 405 nm. This observation confirms the generation of excitons, which undergo hybridization and lead to emission from the polariton branches. However, the optical excitation at 633 nm is lower in energy than the exciton peak at 580 nm. As a result, one might expect the absence of exciton formation and polariton emission. Interestingly, upon excitation at 633 nm, we can still observe the emission from multiple LPBs which suggests modulation of the optical dispersion of the perovskite crystal even without the ability to directly generate excitons. This also suggests, in principle, that sub-bandgap photons can be harvested due to the modified optical dispersion of the perovskite crystal.

We finally investigate the E-P states in Au/RP3, as shown in Figure 1e, which is a more complicated system than Au/RP1 or Au/RP2 due to the presence of layer-edge state (LES) emission at lower energies[30,31] as well as broad emission bands at energies below the bandgap that are found both at the edges of the perovskite crystal and across the whole flake[32,33]. Similar to RP1 and RP2, we observe the emergence of multiple LPBs in reflectance spectroscopy with corresponding PL peaks when using a 405 nm excitation. When excited below the bandgap (610 nm) by the 633 nm laser, the LPB emission is still observed at 680 nm which corresponds to an LPB seen in the reflectance spectrum. To demonstrate that this is LPB emission instead of LES or broadband emission, we first investigated RP3 flakes on Au substrates that are too thin to support cavity modes (<10 nm). The purely excitonic nature of the flakes can be seen in both reflectance and PL from 405 nm excitation spectra since they both have single peaks near the exciton wavelength (Figure S3). However, when excited by light below the bandgap, no emission was observed except for the excitation peak (Figure S3). Therefore, the sub-bandgap emission is not caused by LES or broadband emission since both of these mechanisms are possible in thin flakes.

From the above discussion, we can postulate that the LPB2 emission is a polariton state, not due to LES or broadband emission. However, as discussed above, the emission from LPB2 upon optically pumping below the excitonic energy is similar to RP2 which further suggests modulation of the optical dispersion of the perovskite crystal under these excitation conditions. The exfoliated RP3 flake on the Au substrate was used to record PL data from optical excitation at 405 nm and 633 nm, as shown in Figure 2a. Using the Fourier-plane imaging technique, we recorded the angle-resolved photoluminescence using a 405 nm excitation source at 300 K and 4 K (Figures 2b and 2c). At 300 K, the modes corresponding to LPB1 and LPB2 show a clearly curved dispersion relation, a signature of strong light-matter coupling[34,35]. After cooling down the sample to 4 K, the modes undergo strong coupling. The dispersion from a longer wavelength (660 nm) becomes apparent. The presence of dispersion in the E-k spectra for all wavelengths confirms these states are E-Ps that arise from strong light-matter interaction.

After establishing an understanding of the E-k and PL data, we mapped the perovskite flakes. We spatially resolved them to understand if these emissions were located only at the edges or across the entire flake. As seen in Figure 2d, the excitation at 405 nm shows that both the UPB and HO modes are present across the entire flake. Likewise, the LPB mode is also present across the entire flake. Sub-bandgap excitation with 633 nm shows no emission from the UPB, while the LPB still gives rise to emission across the whole flake (Figure 2e). A similar observation from another sample of RP3 is shown in Figure S5 in the Supporting Information. From the above discussion, we confirm that the exciton-polaritons are the origin of the sub-bandgap emission observed in HOIP flakes and that these emission wavelengths can be tuned by varying the thickness of the flake and introducing additional cavity modes.

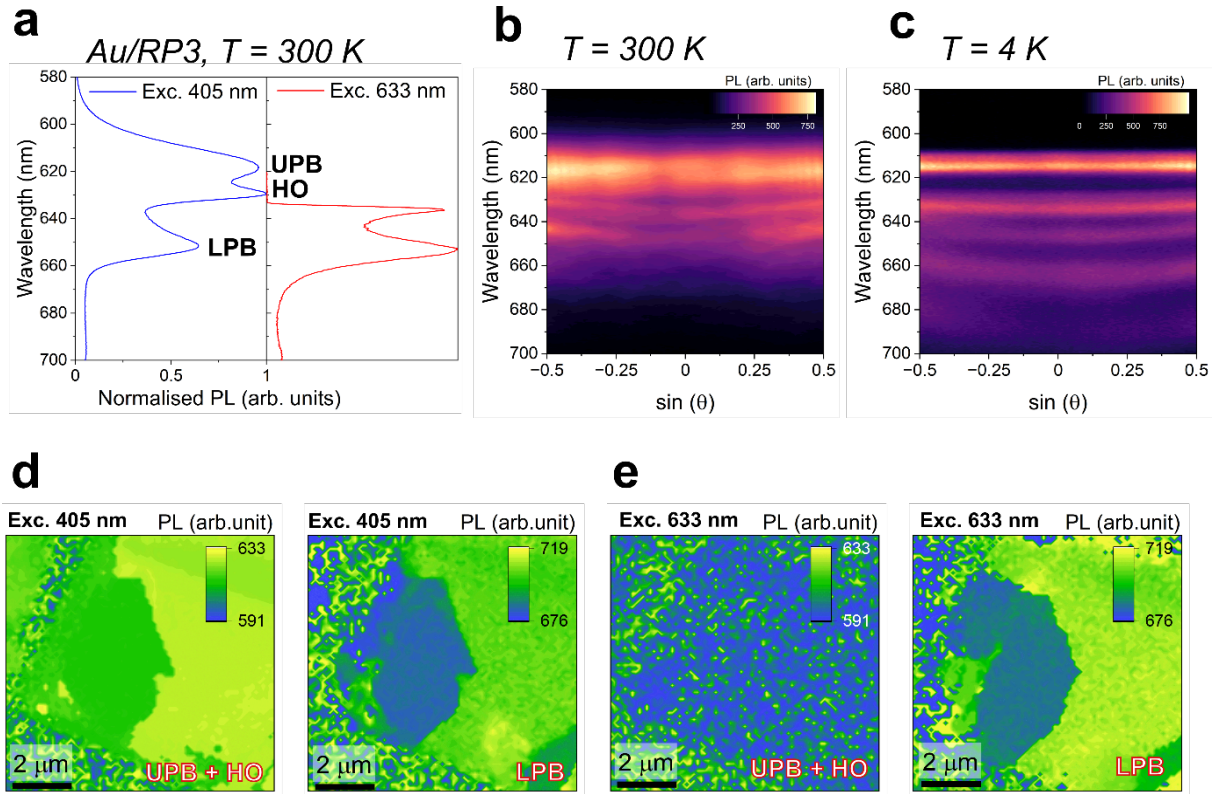

**Figure 2. Sub-bandgap photon absorption/emission across the perovskite flake.** (a) Photoluminescence spectra from the thick RP3 flakes recorded using 405 nm and 633 nm excitation showing exciton-polariton emission. Close match with absorption states in the reflectance spectra at room temperature is shown in Figure 1e. (b) Angle and spectrally resolved photoluminescence at 300 K shows a clear dispersion with the PL peak position matching the steady state spectra in the panel (a). (c) Upon cooling to 4 K, the emission and dispersion are clearly discernable. (d) Using 405 nm excitation wavelength, the PL mapping from the thick perovskite flake selected from the (a) UPB and HO mode and several LPB (LPB1 and LPB2) are shown. All the hybridized modes (UPB, HO, LPB1, LPB2) are observed across the flake. e) For comparison, the PL map using 633 nm excitation wavelength, shows no UPB emission (d), however, the multiple LPB (LPB1 and LPB2) emission modes are observed across the flake (e).

To probe the origin of the emission in the HOIP occurring at a longer wavelength than the E-P, we performed photoluminescence excitation (PLE) spectroscopy on Au/RP3 samples. The ground state absorption spectrum shows multiple absorption bands corresponding to lower polariton branches (Figure 3a). PL corresponding to 631 nm, 650 nm, 686 nm, and 741 nm was recorded in the PLE studies. As seen in Figure 3a, the presence of UPB2, UPB1, and HO modes were observed in the PLE spectra.

This observation confirms the strongly coupled nature of the exciton-polariton, as well as energy transfer from the higher energy UPB to the lower energy LPB states.

Additionally, the presence of UPB2 is clear from the PLE spectrum, which gets suppressed by the density of states in the attenuance spectrum. The absorption spectrum was obtained from transfer-matrix calculations using the optical constants of the perovskites[10]. As reported before, the non-zero value of k above the bandgap results in an overestimated value, while keeping k = 0 underestimates the absorption (see Figure S6 and Figure 3b). A close fit between the experiment and simulation was achieved by fitting the absorption tail with an Urbach function[36], as shown in the inset in Figure 3b. Finally, we also probe the presence of multiple LPBs using transient absorption spectroscopy. Figures 3c and 3d show that multiple LPB branches were observed at the same wavelength, irrespective of the excitation wavelength (450 nm or 650 nm), confirming the energy transfer between the polariton branches. The short lifetime of these states eliminates the possibility of self-trapped exciton defect states[33] or any plexciton states from the Au-HOIP interface[21] since similar lifetimes to what we observed have been reported in glass/HOIP systems[37].

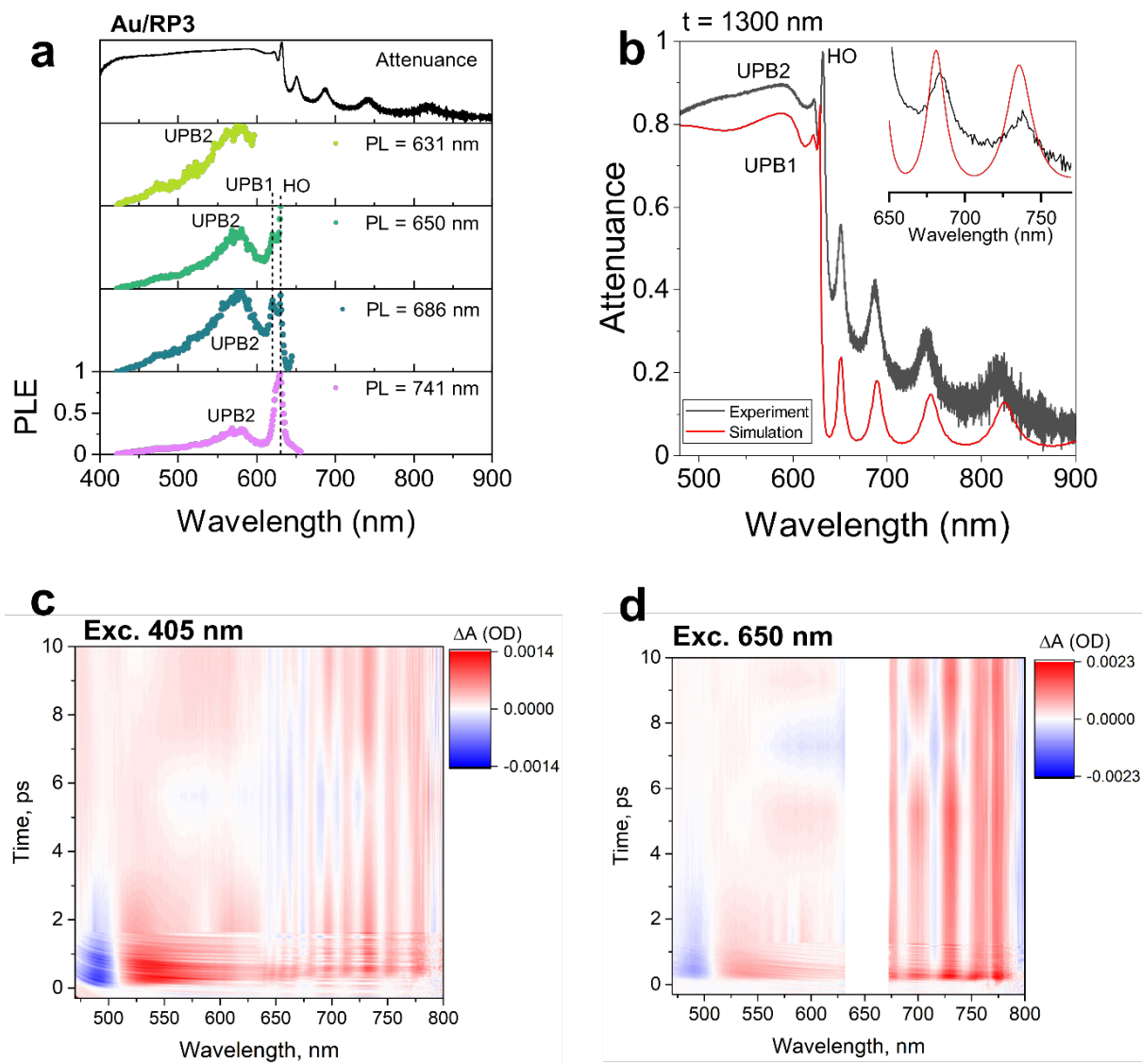

**Figure 3. Steady-state and pump-probe studies confirming sub-bandgap absorption.** (a) Attenuance and photoluminescence excitation spectra for sub-bandgap emissions from RP3 on the Au substrate at room temperature. (b) The experimental attenuance spectrum overlapped with the transfer-matrix simulation data confirming the presence of sub-bandgap absorption states. Time-resolved pump-probe spectra were recorded using (c) 405 nm and (d) 650 nm excitation lasers at 300 K. Cavity mode (>640 nm) absorption emerges in ps time scale and is present even upon pumping below the exciton absorption (605 nm).

To further understand the exciton-polariton dynamics from above-bandgap and sub-bandgap excitation, time-resolved PL (TRPL) was recorded from the Au/RP3 system at 300 K and 4 K. The TRPL decay trace recorded at 300 K for excitation from 470 nm and 640 nm is shown in Figure 4a and 4b, respectively. The bi-exponential decay function is used to deconvolute the decay trace; the resulting fast decay lifetime ($t_1$) and slow decay lifetime ($t_2$) are summarized in Figures 4c and 4d, respectively. We note that the sub-bandgap excitation (640 nm) leads to a longer lifetime than the above bandgap excitation (470 nm). As these hybridized states emerge from strong coupling between the exciton and photon, we hypothesize that these states' exciton and photon fractions may be modified depending on the excitation wavelength. As a result, a higher photon fraction can lead to a longer lifetime (limited by the cavity decay rate) than more exciton fraction (limited by radiative rate). Based on these arguments, we confirm the presence of a different fraction of exciton and photon in these hybrid states. Upon cooling to 4 K, the lifetime of the multiple LPBs upon excitation at 470 nm remains unperturbed compared to the lifetime recorded at 300 K. This behavior was also observed in the HOIP excitons in our previous study[24]. However, the lifetime of the multiple LPBs decreases upon cooling from 300 K to 4 K for excitation at 640 nm. A similar lifetime for excitation at 405 nm confirms the dominance of excitonic characteristics in the multiple LPBs, while the excitation at 640 nm indicated a higher photon fraction.

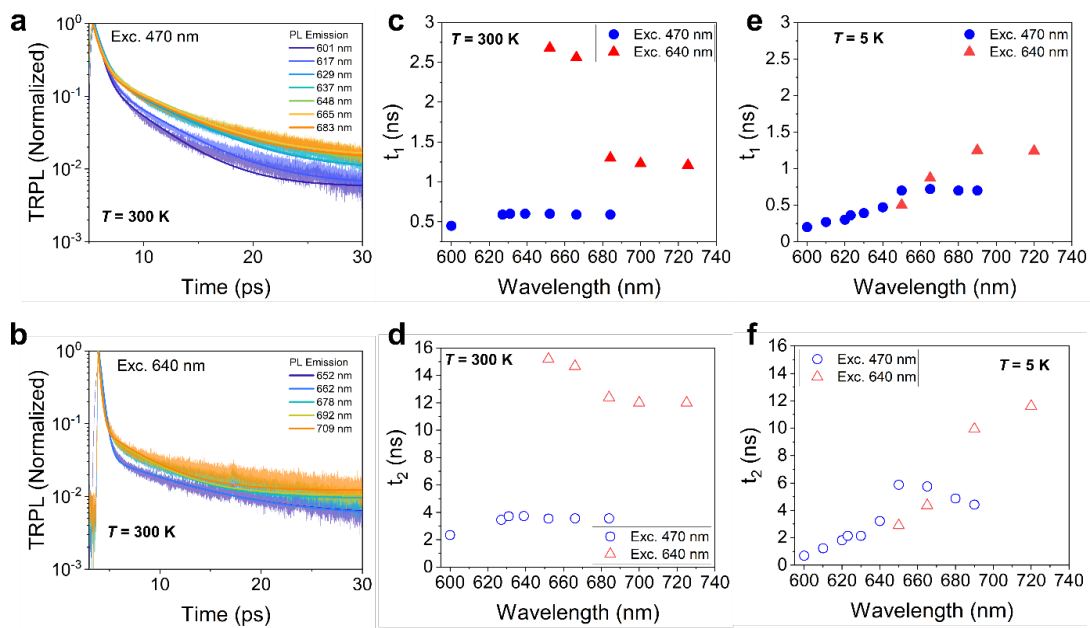

**Figure 4. Time-resolved photoluminescence (TRPL) studies of self-hybridized exciton-polariton states.** TRPL from exciton-polaritons and cavity modes emission

was recorded using 470 nm (a) and 640 nm (b) excitation sources at 300 K. Summary of fast decay component - $t_1$ (c) and slow decay component - $t_2$ (d) from both excitations at 300 K. For comparison, the low-temperature (4 K) exciton-polariton lifetime for $t_1$ (e) and $t_2$ (f).

From the above studies, the mechanism behind the emergence of sub-bandgap absorption and emission is clear. Additionally, the energy transfer between these states is apparent from the PLE and TRPL studies. Further, PL mapping reveals that the emission from all these states are seen across the whole region of the flake suggesting a uniform slab of sample which supports self-hybridized exciton-polaritons in an open-cavity/external cavity free structure. Such a sample presents a suitable opportunity for studying charge transfer from the polaritons to an acceptor state (say, few-layer graphene (FLG) or a metal), which can be fundamentally important for polariton based energy harvesting applications[38]. To investigate this, first reflectance and PL spectroscopy were used to confirm the existence of the lower polariton branch in both absorption/emission in the HOIP flakes, upon which FLG was deterministically transferred to create Au/RP2/FLG and Au/RP3/FLG 2D heterostructures (HS). As seen in the reflectance spectra in Figure 5a, the LPB shows an enhanced absorption after placing this graphene layer while the absorption above the bandgap decreases. To keep the excitation density approximately the same for both cases, we recorded the PL with excitation at 405 nm. At 405 nm, the relative difference in absorption for the HS is 9.2%, while the integrated PL decreases by 80.75% in the same region of the flake (Figure 5b). PL mapping reveals that PL quenching occurs in the HS, but not in the Au/RP2 region (Figure 5c and 5d). Similarly, the RP3/FLG HS shows a similar trend with enhanced absorption and PL quenching (Figure 5e, f). At 405 nm, the relative increase in the absorption for the HS is 8%, while the integrated PL decreases by 64%. The above observations confirm that charge transfer occurs via polaritons at room temperature in an HOIP/graphene HS. This experiment suggests that polaritons can also be used to transport and extract charges from a strongly coupled excitonic medium, opening a new avenue for harvesting sub-bandgap photons for polaritonic photodetectors and photovoltaics.

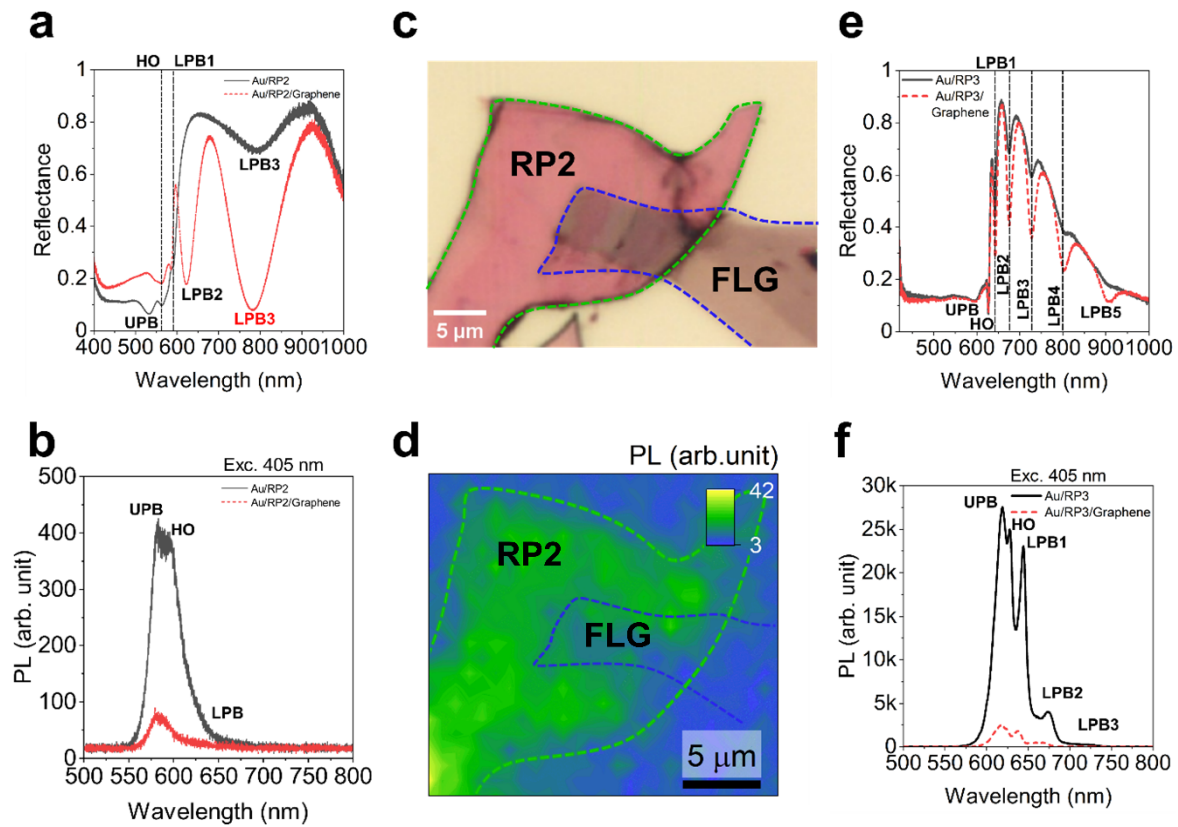

**Figure 5. Charge transfer at the perovskite-few-layer graphene (FLG) heterostructure.** (a) Reflectance spectra show a shift in the exciton-polariton modes and enhanced absorption in the LPB2 mode (780 nm) upon placing few-layer graphene on the RP2 flakes pre-exfoliated on the Au substrate. (b) Exciton-polariton PL quenching from the graphene-covered RP2 flakes. (c) Optical microscopy image of the FLG transferred on RP2. (d) PL map showing polariton emission quenching at the RP2/FLG area. (e,f) FLG on RP3 enhances sub-bandgap absorption of the multiple polariton branches (marked 1 to 5), with polariton emission quenching observed from RP3/FLG heterostructure.

In summary, we have developed an open-cavity system that supports self-hybridized exciton polaritons in 2D HOIP crystals and studied their charge and energy transfer dynamics in detail. The exciton-polariton hybridization modulates the optical dispersion of the perovskite enabling absorption of sub-bandgap photons which also corresponds to sub-bandgap emission. We have further identified that the presence of strong light-matter coupling also supports energy transfer from the higher energy upper polariton branches to multiple lower polariton branches that have lower energies. The exciton and photon fractions in these lower polariton branches can also be modulated using the excitation wavelength, opening new avenues for ultrafast control of polaritonic devices and switches. Finally, by means of polaritonic PL quenching, we have also shown proof of concept that polaritons can induce charge transport and transfer in a strongly-coupled excitonic medium, thereby opening new possibilities for polaritonic photodetectors and photovoltaic devices.

**Methods**

**Sample preparation.** The solution processing method described elsewhere was chosen to obtain single-phase pure 2D perovskite crystals[25–28]. The powders from this method were chosen to obtain the micrometer-thick flakes used in this work. Single crystal perovskite powders were exfoliated onto template-stripped gold substrates using a the tape method well known for exfoliation of layered van der Waals crystals.

**Steady-state optical spectroscopy studies.** Room-temperature reflectance was recorded from Au/perovskite samples in a vacuum environment to avoid perovskite degradation. The absolute reflectance was obtained by normalizing the reflectance from the sample to the reflectance from a silver mirror. For low-temperature measurements, the samples were cooled down to as low as 80 K using liquid nitrogen coolant. The photoluminescence signals from the perovskites were recorded using a continuous-wave excitation source at 405 nm and 633 nm, maintaining a low excitation power to avoid perovskite photobleaching. Both photoluminescence and reflectance spectra were recorded in a Horiba HR Evolution setup by passing the reflected/emitted light through a 600 grooves/mm grating before reaching the charge-coupled detector. Photoluminescence excitation (PLE) spectroscopy was conducted at room temperature to further understand the origin of photoluminescence from the exfoliated flakes. A tunable laser source (SuperK Fianium) was used to scan the excitation wavelength from 420nm to 664nm, while recording the integrated PLE spectra from the Au/perovskite sample for the various photoluminescence signals. The PL signals (631 nm, 650 nm, 686 nm, and 741 nm) were spectrally separated using scombinations of edge pass filters and were recorded using an avalanche photodiode (PDM, micro photon devices). The samples were kept under vacuum ($10^{-5} – 10^{-6}$ torr) during PLE measurments.

**Transient optical spectroscopy studies.** The lifetimes of the exciton-polariton peaks and the sub-bandgap states were probed using time-resolved photoluminescence spectroscopy at room temperature and 4 K. To understand energy transfer between the light-matter coupled states, the photoluminescence signals were recorded from the sample by exposing them to two different laser wavelengths (405 nm and 640 nm) with picosecond excitation and 100 MHz repetition rate, separately. The wavelength-selective lifetime was recorded by passing the photoluminescence signal through two tunable filters and fiber coupling the signal to a single-photon detector with a PicoQuant HydraHarp 400 timing box. The data collected from the streak camera was used to analyze the lifetime of the different states with each of the two different wavelength excitations (450nm and 650nm).

Pump-probe spectroscopy studies are performed on the Au/perovskite samples at the ambient conditions using a single-wavelength laser pulse as a pump and a broadband white-light continuum pulse as a probe. The pump-probe signal is recorded as a transient reflectivity change for the probe pulse as a function of time delay between the pump and probe pulses. The pump pulse is generated using a Coherent Astrella Ti:Sapphire amplified system centered at 800 nm and an optical parametric amplifier (OPA). The broadband probe beam (400 nm - 1000 nm spectral range) is generated by focusing the output of the amplifier on a Sapphire crystal. The pump beam is at normal incidence to the sample while the probe beam is deviated from normal by a small angle to record the reflected signal.

**2D perovskite/FLG heterostructures and perovskite encapsulation:** Charge transfer from the polariton states was investigated by placing few-layer graphene (FLG) on the perovskite, which had already been exfoliated on the Au substrate. The dry transfer of the FLG onto the perovskite was performed at room temperature inside a nitrogen-filled glovebox to suppress perovskite degradation. Additionally, encapsulation of the perovskites was investigated by spin coating the polystyrene solution at different rpm. Nevertheless, the reflectance and photoluminescence were recorded in a vacuum using the Horiba HR Evolution.

**Atomic force microscopy:** The thicknesses of the perovskite flakes exfoliated on the Au substrates were estimated using OmegaScope Smart SPM (AIST).

**Simulation:** Transfer-matrix calculation[39] was used to explain the sub-bandgap peaks emerging below the exciton absorption as a result of strong coupling[24]. The optical constants for the perovskites used here are reported elsewhere[28]. The room-temperature experimental reflectance data from the sample was compared against the simulation data to identify the value of k below the exciton bandgap.

**Author Contributions**

S.B.A. and D.J. conceived the idea. S.B.A designed and executed the project through collaborations initiated by D.J.. J.L. performed the transfer matrix calculations to estimate the absorption bands from the perovskite/Au interface. C.E.S and C.M performed the TRPL and E-k measurements, supervised by J.R.H.. C.L. performed transient absorption measurements under the supervision of H.H.. J.H., H.Z., and A.T. performed PLE measurements under the supervision of A.M. The results from PLE and E-k measurements were discussed with T.D., J.S and A.M., respectively. F.C. and K.L. performed the steady-state optical measurements for perovskite-graphene and perovskite-polystyrene structures.

## Acknowledgments

D.J. acknowledges primary support for this work by the U.S. Army Research Office under contract number W911NF-19-1-0109 and the Asian Office of Aerospace Research and Development of the Air Force Office of Scientific Research (AFOSR) FA2386-20-1-4074. S.B.A. gratefully acknowledges funding received from the Swiss National Science Foundation (SNSF) under the Early Postdoc Mobility grant (P2ELP2_187977) for this work. C.M. is supported by an NSF-AFRL Intern Program. The experiments were carried out at the Singh Center for Nanotechnology at the University of Pennsylvania, which is supported by the National Science Foundation (N.S.F.) National Nanotechnology Coordinated Infrastructure Program grant NNCI-1542153. D.J. and K.L acknowledge the NSF REU SUNFEST program under Grant No. 1950720, to support the stay of K.L. at the University of Pennsylvania. The research performed by C.E.S. at the Air Force Research Laboratory was supported by contract award FA807518D0015.  J.R.H. acknowledges support from the Air Force

Office of Scientific Research (Program Manager Dr. Gernot Pomrenke) under award number FA9550-20RYCOR059. T.D. and P.J.S. gratefully acknowledge support from Programmable Quantum Materials, an Energy Frontier Research Center funded by the U.S. Department of Energy (DOE), Office of Science, Basic Energy Sciences (BES), under award DE-SC0019443.

# Supporting Information

**Charge and Energy Transfer Dynamics of Hybridized Exciton-Polaritons in 2D Halide Perovskites**

*Surendra B. Anantharaman[1,*], Jason Lynch[1], Christopher E. Stevens[2,3], Christopher Munley[4], Chentao Li[5], Jin Hou[6,7], Hao Zhang[6,7], Andrew Torma[6,7], Thomas Darlington[8], Frank Coen[1], Kevin Li[1], Arka Majumdar[4,9], P. James Schuck[8], Aditya Mohite[5,6], Hayk Harutyunyan[5], Joshua R. Hendrickson[3], Deep Jariwala[1,*]*

[1] Department of Electrical and Systems Engineering, University of Pennsylvania, Philadelphia, Pennsylvania 19104, United States

[2] KBR Inc., Beavercreek, Ohio, 45431, United States

[3] Air Force Research Laboratory, Sensors Directorate, Wright-Patterson Air Force Base, Ohio 45433, United States

[4] Department of Physics, University of Washington, Seattle, Washington 98195, United States

[5] Department of Physics, Emory University, Atlanta, Georgia 30322, United States

[6] Department of Chemical and Biomolecular Engineering, Rice University, Houston, Texas 77005, United States

[7] Department of Materials Science and Nanoengineering, Rice University, Houston, Texas 77005, United States

[8] Department of Mechanical Engineering, Columbia University, New York, New York 10027, United States

[9] Department of Electrical and Computer Engineering, University of Washington, Seattle, Washington 98195, United States

Corresponding authors: surendra.anantharaman@gmail.com , dmj@seas.upenn.edu

**Enhancing environmental and excitation stability of 2D HOIP cyrstals supporting self hybridized exciton-polaritons**

The higher-order (HO) modes emerging from the strong coupling of exciton-polaritons can also serve as an optical sensor owing to the high Q and narrow linewidths that they possess which are extremely sensitive to the surrounding dielectric medium (See figure S6). Monitoring perovskite quality in-situ during optical measurements under high photon flux densities by shifting the excitonic peak is less sensitive due to broad linewidth (15 nm). Here we used the HO mode with ultranarrow linewidth (<3 nm) to monitor the perovskite degradation in ambient atmosphere by comparing two samples - pristine Au/RP2 flakes and polystyrene encapsulated Au/RP2 flakes. As seen in Figure S1, the pristine samples showed a drastic degradation in HO mode by 67% within 24 h. However, the encapsulated sample showed no change after 150 h exposure to ambient atmosphere.

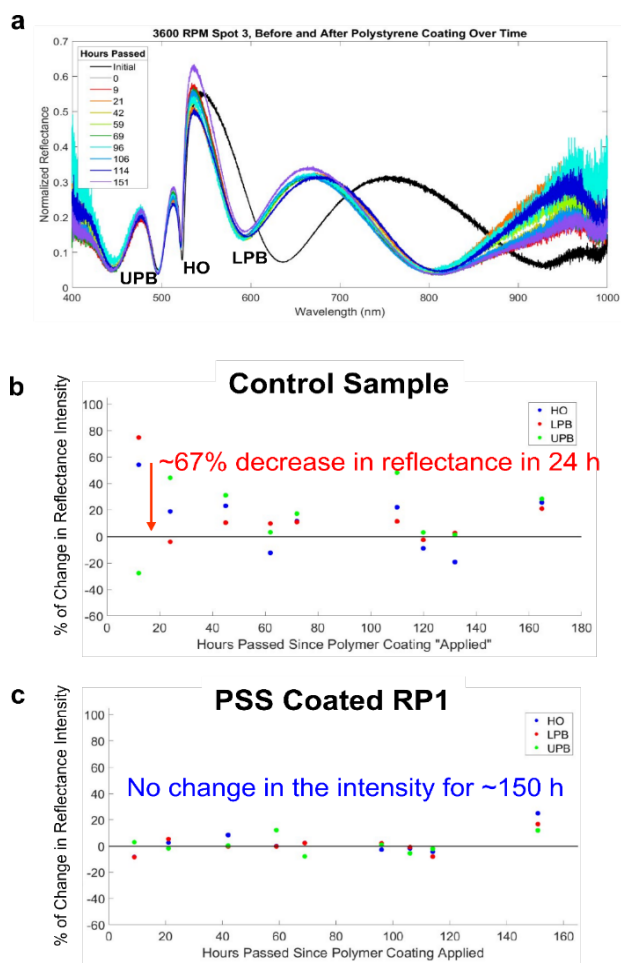

**Figure S1. RP1 encapsulated in Polystyrene and HO mode sensor.** (a) RP1 exfoliated on Au substrate forming HO mode was used as an optical sensor to monitor perovskite degration. (b) Unencapsulated RP1 shows a 67% drop in the perovskite absorption for the HO mode. (c) polystyrene encapsulated RP1 shows no change in the intensity for 150 h in ambient atmosphere, thereby confirming the enhanced stability at room temperature and validating the HO mode as an optical sensor.

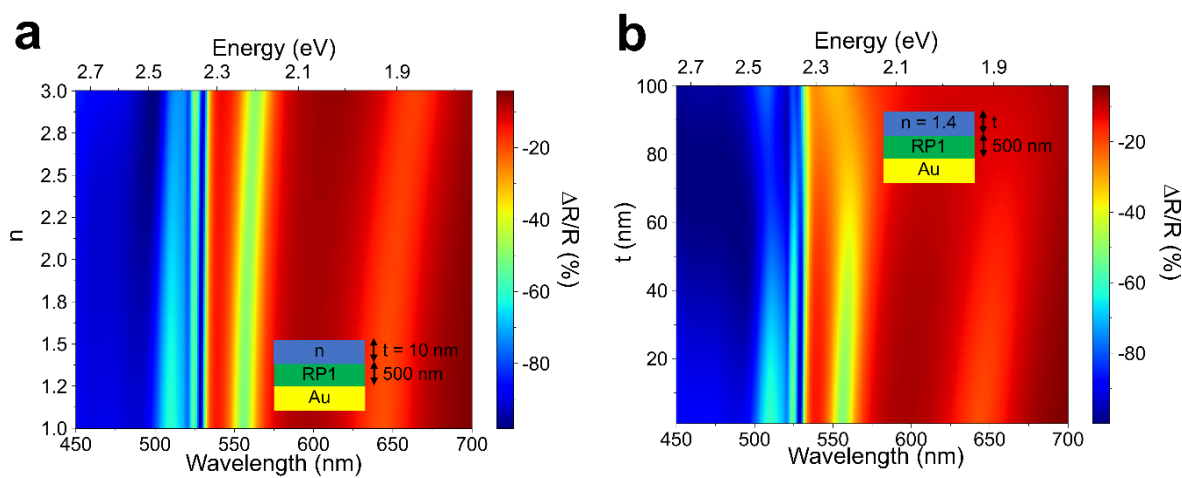

**Figure S2. Sensing using RP1** The normalized change in reflectance for (a) a film of varying refractive index, n, that is 10 nm thick on top of 500 nm of RP1 and an Au substrate, and (b) a film of with a refractive index of 1.4 and varying thickness, t, on top of 500 nm of RP1 and an Au substrate. A refractive index of 1.4 was chosen for (b) since it is comparable to most organic molecules.

**Table S1.** Q-factor of the experimental UPB, HO, and LPB in the thick film (>500 nm) as well as uncoupled excitons in thin films perovskites. The UPB and LPB are the two modes closest to the HO. The Q-factor for the UPB, HO, and LPB was calculated using the reflectance spectra. The Q-factor for the uncoupled exciton is from the fit parameters of our previous work[1].

|  | UPB | HO | LPB | Exciton |
|---|---|---|---|---|
| RP1 | 12.0 | 131.6 | 78.6 | 54.0 |
| RP2 | 8.6 | 101.4 | 52.7 | 39.4 |
| RP3 | 7.3 | 82.3 | 38.0 | 33.4 |

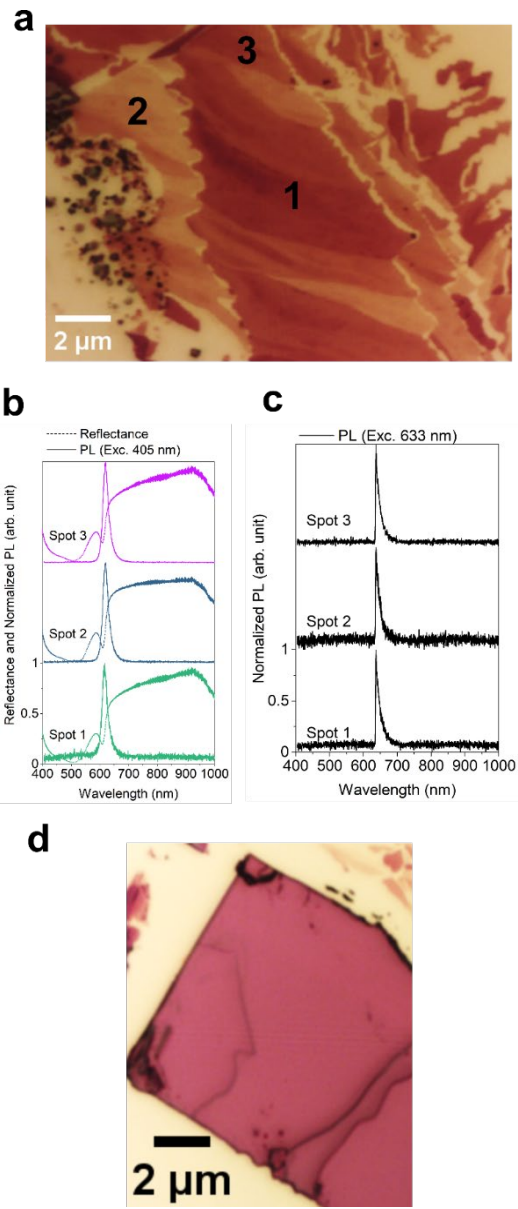

**Figure S3. Exciton emission in RP3 without layer-edge state (LES) emission.** a) Optical image of an exfoliated thin RP3 flakes on the Au substrate with three different thicknesses marked as 1-3. Photoluminescence spectra were recorded at room temperature from spots 1-3 with 405 nm (b), and 633 nm (c) pump laser. No LES emission was observed from thin RP3 flakes. In panel c, the sharp rise in the PL intensity corresponds to the cut-off from the 633 nm laser. d) Optical image of the thick RP3 flake exfoliated on the Au substrate.

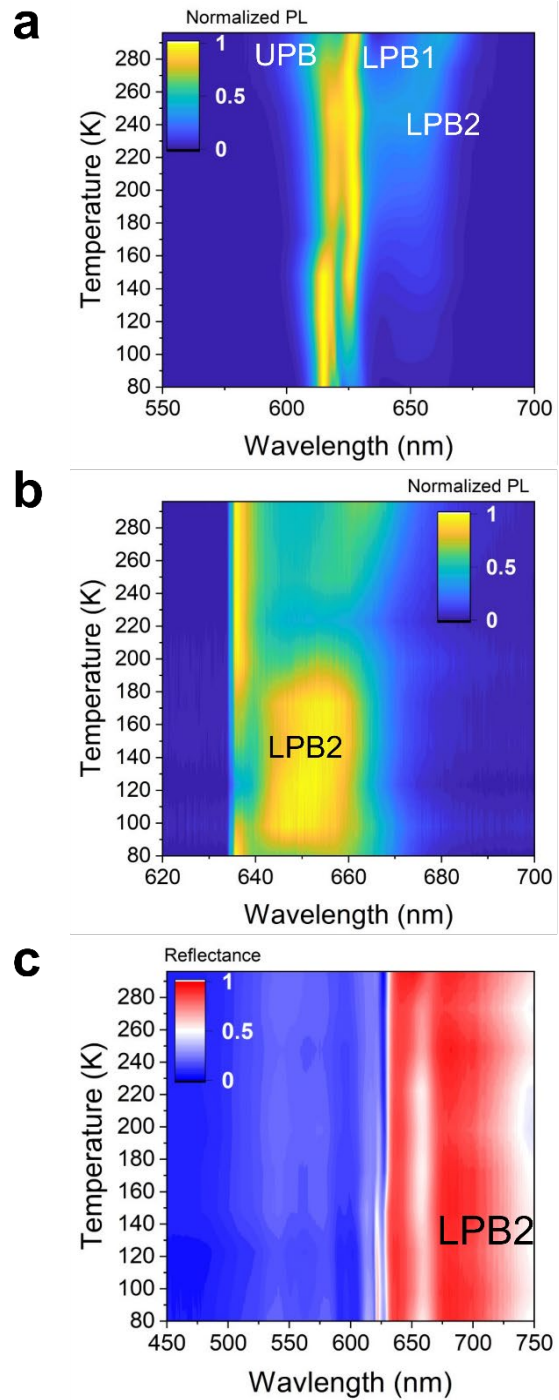

**Figure S4. Temperature-dependent exciton-polariton dynamics in RP3 on Au.** The normalized PL for excitation at 405 nm (a) and 633 nm (b), shows the presence of LPB2 emission at low temperature. Increase in LPB2 absorption at low temperature is clear from the reflectance studies (c).

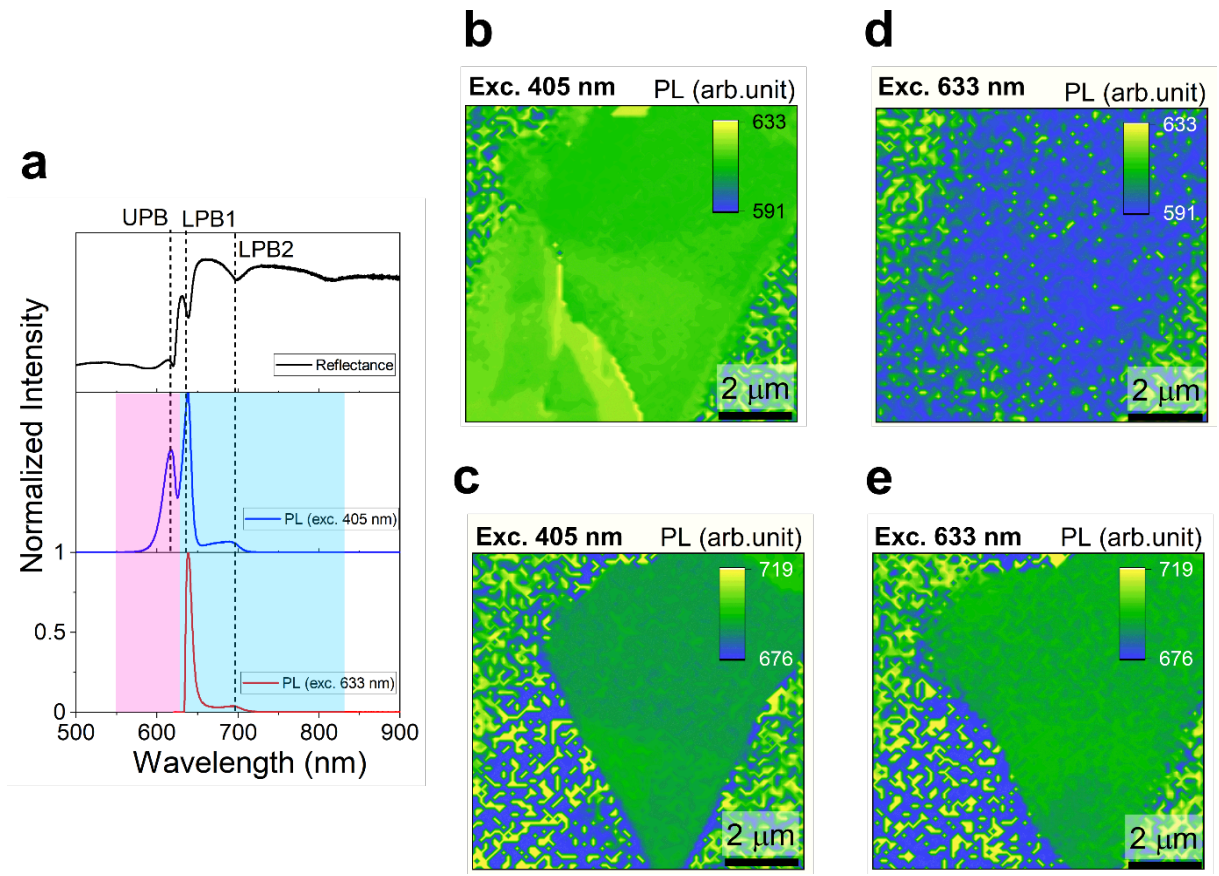

**Figure S5. Room-temperature PL mapping from exciton-polaritons in RP3 exfoliated on the Au substrate.** (a) Reflectance, PL from 405 nm and 633 nm excitation wavelength from RP3 exfoliated on Au. PL mapping for 405 nm excitation showing the presence of UPB and HO (b) and multiple LPB modes (LPB1 and LPB2) (c). PL mapping for 633 nm excitation showing the absence of UPB and HO (d) and presence of multiple LPB modes (LPB1 and LPB2) (e).

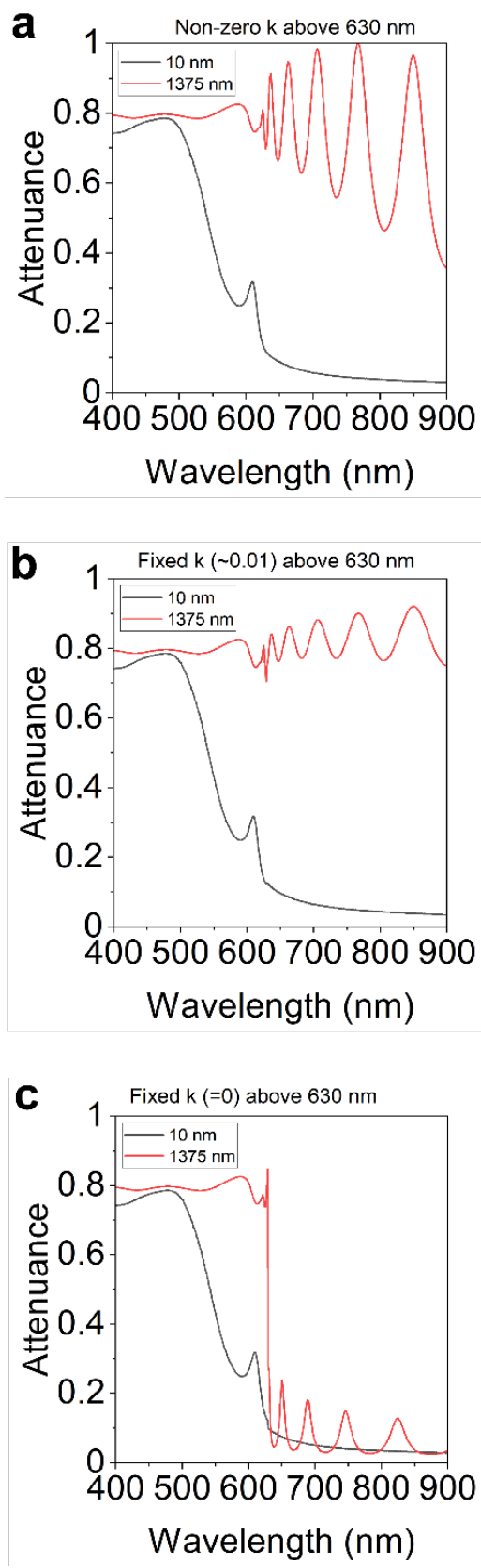

**Figure S6. Attenuance spectra with varying k above bandgap for RP3.** The attenuance spectra calculated for thin flakes (10 nm) and thick flakes (1375 nm) for a non-zero k (a), k = 0.01 (b) and k = 0 (c).